\documentclass[doublecol,amsmath,amssymb]{epl2}

\usepackage{epsf}
\usepackage{rotating}
\usepackage{amsmath,amsfonts}
\usepackage[dvips]{psfrag}

\usepackage{graphicx}
\usepackage{dcolumn}
\usepackage{bm}

\def\hat{\widehat}
\def\tilde{\widetilde}

\def\pmb#1{\setbox0=\hbox{#1}%
  \kern-.025em\copy0\kern-\wd0
  \kern.05em\copy0\kern-\wd0
  \kern-.025em\raise.0433em\box0}

\def\bfnabla{\mbox{\boldmath $\nabla$}}

\def\bfx{\mbox{\bf x}}

\def\bfu{\mbox{\bf u}}
\def\bfB{\mbox{\bf B}}
\def\calS{\mbox{$\cal S$}}
\def\calU{\mbox{$\cal U$}}
\def\calL{\mbox{$\cal L$}}
\def\tauU{\tau_{\mbox{\tiny $\cal U$}}}
\def\calLU{\mbox{$\cal L_{\mbox{\tiny $\cal U$}}$}}
\def\dd{{\rm d}}
\def\rme{{\rm e}}

\def\dR{{\rm d}}

\def\LC{{\cal L}}

\def\RC{{\cal R}}

                 \def \etal{{\it et al.\ }}

\newcount\eqnno
\newcount\secno   
\newcount\subno   

\def\section#1
   {\vskip0pt plus.1\vsize\penalty-250
    \vskip0pt plus-.1\vsize\vskip18pt plus9pt minus6pt
    \subno=0 \eqnno=0 
    \global\advance\secno by 1
   \centerline{\bf\the\secno. #1}
    \medskip}

\def\subsection#1
   {\vskip-\lastskip
        \vskip15pt plus4pt minus4pt
    \bigbreak
    \global\advance\subno by 1
    \centerline{\it \the\secno.\the\subno. #1}\smallskip}

\def\sen{\the\secno}

\def\eqn#1{\global\advance\eqnno by 1
           \eqno(\sen.\the\eqnno)
           \expandafter \xdef\csname #1\endcsname
           {\sen.\the\eqnno}\relax }
\def\eqnp#1#2{\global\advance\eqnno by 1
           \eqno(\sen.\the\eqnno\hbox{#2})
          \expandafter \xdef\csname #1\endcsname
           {\sen.\the\eqnno}\relax }
\def\eqnpr#1{\eqno(\sen.\the\eqnno\hbox{#1})}
\def\eqnn#1{\global\advance\eqnno by 1
           (\sen.\the\eqnno)
           \expandafter \xdef\csname #1\endcsname
           {\sen.\the\eqnno}\relax }
\def\eqnm#1#2{\global\advance\eqnno by 1
           (\sen.\the\eqnno\hbox{#2})
           \expandafter \xdef\csname #1\endcsname
           {\sen.\the\eqnno}\relax }
\def\eqnr#1{(\sen.\the\eqnno\hbox{#1})}

\catcode`\£=\active \gdef£{\setbox0=\hbox{0}\hbox to\wd0{}}
\def\Reym{\mbox{\it Rm}}  
\newsavebox{\thalfbox}
\sbox{\thalfbox}{$\textstyle\frac{1}{2}$}

\newsavebox{\shalfbox}
\sbox{\shalfbox}{$\scriptstyle\frac{1}{2}$}

\newsavebox{\squartbox}
\sbox{\squartbox}{$\frac{1}{4}$} 

\newsavebox{\etbox}
\sbox{\etbox}{\boldmath$\eta$}

\newsavebox{\astrutbox}
\sbox{\astrutbox}{\rule[-5pt]{0pt}{20pt}}

\title{Time scales separation for dynamo action} 
\author{E. Dormy\inst{1,3} \and D. G\'erard-Varet \inst{2,3}}
\shortauthor{E. Dormy \& D. G\'erard-Varet}
\institute{                    
  \inst{1} MAG (Ecole Normale Sup\'erieure / Institut de Physique du
  Globe de Paris),
D\'epartement de Physique, LRA,\\ Ecole Normale Sup\'erieure --
24, rue Lhomond, 75231 Paris Cedex 05, France.\\
  \inst{2} D\'epartement de Math\'ematiques et Applications,\\
Ecole Normale Sup\'erieure -- 
 45, rue d'Ulm, 75230 Paris Cedex 05, France.\\
  \inst{3} C.N.R.S., France.
}

\pacs{47.65.-d}{Magnetohydrodynamics}
\pacs{47.65.Md}{Plasma dynamos}
\pacs{47.20.-k}{Flow instabilities}

\abstract{The study of dynamo action in astrophysical objects classically
involves two timescales: the slow diffusive one and the fast advective
one. 
We investigate the possibility of field amplification on an intermediate
timescale associated with time dependent modulations of the flow.
We consider a simple steady configuration for which dynamo action is
not realised. We study the effect of time dependent perturbations of 
the flow. We show that some vanishing 
low frequency perturbations can yield exponential 
growth of the magnetic field on the typical time scale of oscillation. 
The dynamo mechanism relies here on a parametric instability associated 
with transient amplification by shear flows. Consequences on natural
dynamos are discussed.
}

\begin{document}

\maketitle

\noindent {\bf Introduction.}\! --
Most astrophysical objects exhibit magnetic fields. This is the case of
planets, stars, as well as galaxies. The dynamo instability, characterised by
an exponential growth of the magnetic field in a magnetohydrodynamic flow, 
is expected to account for these fields.
Among dynamo flows, a vast majority provides finite growth rate on the
diffusion time only (known as ``slow dynamos'').
They are therefore not suitable for large astrophysical objects (stars and
galaxies), for
which the associated magnetic Reynolds number (measuring the relative
strength of the induction to the diffusive term) is huge. 
This is the primary motivation for the investigation of fast dynamos,
characterised by finite  growth rate on the advective time scale rather
than the diffusive time  scale
(the later being sometimes larger than the age of the universe). 

While significant progress has been achieved in the case of moderate 
values of the magnetic Reynolds number \cite{Glatz,Christ},
the fast dynamo limit has been tackled yet only with a few
analytic flows \cite{Fast1,Fast2}, still remote from realistic configurations.
Criteria to assess dynamo action (other than direct integration) are still 
missing. Most simple flows (such as uniform shear) fail to be dynamos and only
yield transient amplification of the field, followed by ohmic decay.

We discuss the possibility of a third class of dynamo, for which
the field grows on an intermediate timescale associated to flow 
modulations (for example due to a traveling wave). The efficiency of some
time dependent flows in the dynamo process has been recognised in various 
studies 
{
\cite{Backus,Soward,Soward2,Fast2,Gog}. 
These studies focused on leading order time dependence of the flow 
(either analytical, or in the form of an heteroclinic cycle), whereas 
we want to study the effects of vanishing perturbations on steady flows.
We stress that the impact of small fluctuations on dynamo flows has
already been addressed in other contexts.
Gog \etal \cite{Gog} considered the action of
a (random) noise on the dynamo properties of time dependent flows
provided by heteroclinic cycles. 
P\'etr\'elis \& Fauve \cite{PF} and  Peyrot \etal \cite{Peyrot} 
investigated the effects of
fluctuations on the dynamo threshold (respectively for the G.O. Roberts 
and the Ponomarenko flows).}

The magnetic field evolution in a conducting fluid is governed by the 
induction equation
\begin{equation}
\partial _t \bfB = \bfnabla \times (\bfu \times \bfB) + \eta \, \Delta \bfB
 \, .
\end{equation}
The rhs operator is not self-adjoint \cite{induction,Farrell,induction2,induction3}
and can therefore provide algebraic 
growth of an initial perturbation \cite{Schmid}.
This transient magnetic amplification is typically observed in 
a uniform shear (as examplified by the $\Omega$--effect).
We will see how this effect can be used to yield 
dynamo instability under appropriate time dependent disturbances.

\noindent {\bf Modelling.}\! --
We focus on a very simple setting, retaining only the essential 
physical 
characteristics of the above system (the transient amplification).
Let us consider a two-dimensional flow $\bfu =(u_x(y,t),\,0,\,0)^t$ 
and further assume an initial magnetic field of the form $\bfB_0 = (B_x \,
\cos(z), \,B_y \, \cos(z), \, 0)^t$ independent of the position $\bfx$. 
The field depends only on 
the space coordinate $z$, which provides a typical
length scale $L$.  

We assume that the flow consists of a uniform and constant shear
$\partial_y u_x(y,t) \equiv \calS$~. 
The evolution of $B_x$ and $B_y$ in time is then governed by
\begin{equation}
\partial_t 
\begin{pmatrix}
B_x\\B_y
\end{pmatrix}
=  
\begin{pmatrix}
- \eta & \calS \\
0      & - \eta
\end{pmatrix}
\begin{pmatrix}
B_x\\B_y
\end{pmatrix} 
=
\calL
\begin{pmatrix}
B_x\\B_y
\end{pmatrix} 
\, .
\label{goveq}
\end{equation}
The time variation of the field strength is measured by
\begin{align}
||\bfB||^2 &=
\frac{1}{2 \pi}\,\int_0^{2\pi} B_x^2 +B_y^2 \, \dd z \nonumber \\&=
\frac{1}{2}\, \rme^{-2\eta t}
\left[ \left(B_x(0) + \calS \, t \, B_y(0) \right) ^2
+ B_y(0) ^2 \right] \, ,
\label{transientgrowth}
\end{align}
which clearly highlights the transient algebraic amplification for
an initial seed field along $B_y$.

We will now discuss the effect of a time dependent perturbation of the 
flow. Two typical time scales can be identified
in the original problem
\eqref{goveq}.
The advective 
time scale $\tauU = L/\calU = \calS^{-1}$, and the diffusive time scale 
$\tau_\eta = L^2/\eta$.
As we focus here on the 
limit of large magnetic Reynolds numbers
 (relevant to astrophysical dynamos), 
we shall  preferably use the advective time scale.
The system can be rescaled accordingly
using $\varepsilon = \eta / \calU  L$ 
(the inverse of the magnetic Reynolds number), and
\begin{equation}
\calLU=\begin{pmatrix}
- \varepsilon & 1 \\
0      & - \varepsilon
\end{pmatrix} \, .
\end{equation}

We will now consider small amplitude, time-periodic, perturbations
of the flow $\bfu$. These could for example represent the effect of a
traveling wave. For simplicity,  we assume that the 
perturbation modifies the direction of the flow but not its amplitude.
The pulsation is described by a sinusoidal wave of 
amplitude $\alpha$ and frequency $\beta$.
The new velocity field can therefore be expressed as 
$\bfu^{\rm new}=\RC(\theta(t)) \, \bfu$~,
where $\RC(\theta)$ is the $2\!\times\!2$ rotation matrix of angle
$\theta$, and $\theta(t) = \alpha \, \sin (\beta t)$. 
The evolution of the magnetic field $\bfB$ is then governed by
the linear operator
\begin{equation}
\LC(t) = \RC(\theta(t)) \,\, \calLU \,\,\RC(\theta(t))^{-1} \, .
\label{system_puls}
\end{equation} 
Note that we have just introduced a third time scale
in the problem, 
$\tau_\beta \sim L / (\beta \, \calU)$.

We will investigate the field evolution and the possibility of
dynamo instability, yet the system was made
very simple: the field is 
independent of $x$ and $y$, and the flow is planar ($u_z \equiv 0$).
It is then natural to ponder the feasibility of dynamo action. In fact, the
simplicity of the model will also prevent the application of 
the ``two-dimensional field'' and ``planar velocity'' anti-dynamo
theorems~\cite{Zeldovich,Zeldovich2}. As the field is independent of $x$ and $y$,
an expression in the form of a streamfunction would necessarily yield 
unbounded values: the field considered here
does not decay to infinity. The 
above anti-dynamo theorems therefore do not apply to this setup.

Rewriting this system in the rotating frame relative to which the flow
is steady yields 
\begin{equation}
\partial_{t}
\begin{pmatrix}
B_x\\B_y
\end{pmatrix}
=  
\begin{pmatrix}
-\varepsilon & 1 +  \dot{\theta}(t)   \\
-\dot{\theta}(t) &-\varepsilon 
\end{pmatrix}
\begin{pmatrix}
B_x\\B_y
\end{pmatrix} 
\, .
\label{pulsations}
\end{equation} 
This problem can be tackled through the change of unknown
$\bfB' = \rme^{\varepsilon t}\, \bfB$~, 
and a rescaling in time to the $\tau_\beta$ unit of time, 
$t' = \beta t$ which leads to
\begin{equation}
\partial_{t'}
\begin{pmatrix}
B_x'\\B_y'
\end{pmatrix}
=  
\begin{pmatrix}
0 & {1}/\beta + \alpha \cos t' \\
-\alpha  \cos  t' & 0
\end{pmatrix}
\begin{pmatrix}
B_x'\\B_y'
\end{pmatrix} 
\, .
\label{system}
\end{equation}

Let us stress that, for simplicity, we have introduced this system 
in a very restricted setup, however such formulations can 
be extended to local behaviour in more complex flows,
for instance near stagnation points of pulsed Beltrami waves 
\cite{Soward2}.

\noindent {\bf Floquet analysis.}\! --
The stability of this system can be investigated depending on the values 
of $\alpha,\beta$.
We will in particular focus on the limit of small $\alpha$, i.e.
the limit of infinitesimal perturbations of the flow.
This problem fits the framework of the Floquet analysis, and the stability
is determined by the logarithm of the resolvent matrix eigenvalues. 
{Numerical computation of these eigenvalues are reported Figure~1.
For a perturbation of a definite temporal form $\beta$, finite
growth rate can only be achieved for finite amplitude modulations: there
exists a threshold in the amplitude $\alpha$ below which $\sigma'$
vanishes (and the unfiltered variable $\bfB$ decays). 
This reflects the impossibility of fast dynamo action (i.e. growth on the 
advective time scale) with the chosen flow.
However, in the limit of timescales separation, that is
\begin{equation} 
\tau_U \ll \tau_\beta \ll \tau_\eta \, ,
\end{equation} 
which yields
$\beta \rightarrow 0$,
exponential growth can occur on the $\tau_\beta$ time scale, as
revealed by the finite growth rate $\sigma '$ at any value 
of $\alpha$ for small enough $\beta$. }
Tiny disturbances can yield 
exponentially growing solutions on a typical time scale provided by the
perturbation (i.e.~$\tau_\beta$). 
This yields the notion of time scales separation. 
Remarkably, the growthrate does not always vanish
in the limit of small $\alpha$ (i.e.~infinitesimal perturbation).

\begin{figure}
\onefigure{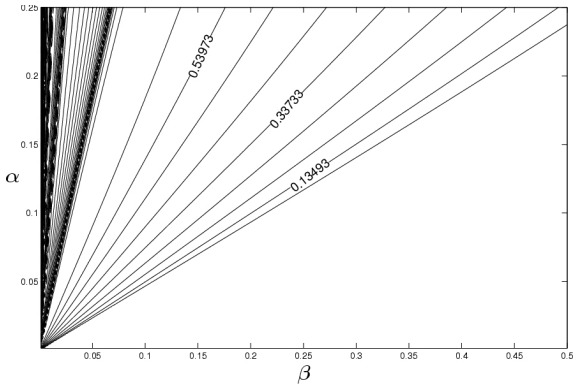}
\caption{Growth rate $\sigma'$  (in the $\tau_\beta$ time scale) 
 as a function of  $\alpha$ and $\beta$.}
\end{figure}

Figure 1 reveals a scaling of the form $\alpha \propto \beta$,
which points to the use of $\gamma=\alpha/\beta$.
We investigate on Figure~2a the growth rate of \eqref{system}
as a function of 
($\alpha, \, \gamma$). The system has an intricate behavior, with large
regions of instability, even in the limit $\alpha \rightarrow 0$~, 
{at finite $\gamma$,}
separated by narrower {neutrally stable bands. Finite growthrate
  on the $\tau_\beta$ time scale for large $\gamma$ (small $\beta$ at fixed
$\alpha$) is even clearer than on Figure~1.}
This behavior is reminiscent
of the parametric instability that develops in forced mechanical systems.
A typical example is provided by the Mathieu equation~\cite{Morse}
(governing a vertically excited pendulum), 
\begin{equation}
\ddot \theta + \left(\tilde{\alpha} + \gamma \cos t \right) \, \theta =0 \, .
\label{eq:Mathieu}
\end{equation}

\begin{figure}
\onefigure{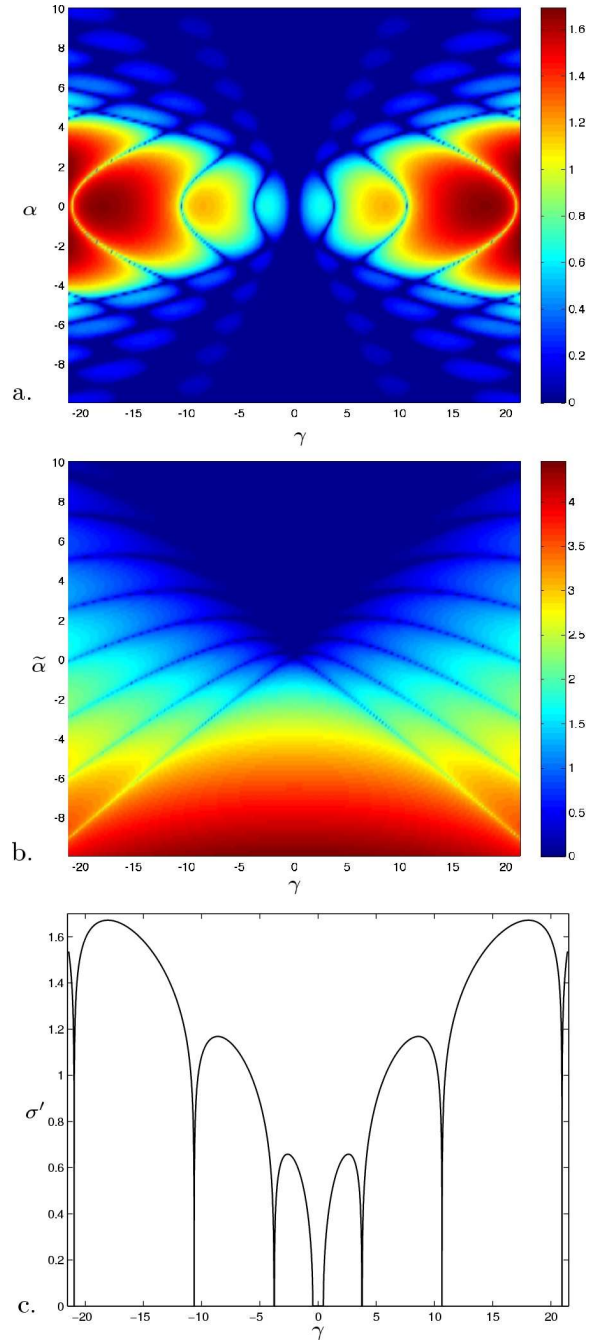}
\caption{(Color online) a. Growth rate $\sigma'$
represented in the $(\gamma,\alpha)$--plane.
b. Growth rate for the Mathieu equation in the $(\gamma,\tilde \alpha)$--plane.
c. Both solutions are identical when $\alpha = \tilde \alpha = 0$.}
\end{figure}

The classical stability diagram for this system is represented on Figure~2b.
Note that  system \eqref{system} can be rewritten
in the scalar form
\begin{equation}
\ddot B_x' 
+ \frac{\alpha^2 \, \sin t}{\gamma + \alpha^2 \, \cos t}\,  \dot B_x'
+ \cos t \, (\gamma + \alpha^2 \, \cos t) \, B_x' = 0 \, .
\end{equation}
It is worth stressing that this system does not reduce in an appropriate
limit to the unforced pendulum, i.e.~\mbox{$\gamma = 0$} in \eqref{eq:Mathieu}.
An equivalence between both problems can, however, be established in the limit 
$\alpha, \tilde{\alpha} \rightarrow 0$ (which corresponds
to vanishing gravity for the pendulum).
This is highlighted on Figure~2c which presents cross sections
of Figures~2a and 2b for $\alpha=0$ and $\tilde \alpha=0$ respectively.

The growth rate
$\sigma'(\alpha,\gamma)$ seems unbounded
as $\gamma \rightarrow \infty$, but this is only
an effect of the chosen 
time scale $\tau_\beta$, which
becomes increasingly long compared to $\tauU$ 
in this limit. Translating to the advective time scale $\tauU$, 
the growth rate is given by
$\sigma(\alpha, \gamma)=\beta \sigma'(\alpha, \gamma) = |\alpha|
\sigma'(\alpha, \gamma) / |\gamma|$. 
The quantity $\sigma''\equiv\sigma'(\alpha, \gamma) /|\gamma|$ is
represented on Figure~3, which clearly shows that $\sigma''$ is
uniformly bounded for all $(\alpha,\gamma)$ and reaches positive values.
This validates the scaling $\alpha \sim \beta$, the optimum 
growth rate
is achieved for $\gamma\approx0.786$, and behaves like  $\sigma\approx0.454\,
\alpha$~for small $\alpha$. 

One should keep in mind that ohmic decay has been 
filtered out from \eqref{pulsations} to \eqref{system}: exponential
growth therefore requires $\sigma > \varepsilon$ and thus
$\alpha > \varepsilon$. 
This corresponds to ensuring that a modified magnetic Reynolds number 
constructed as $\tau_\eta / \tau_\beta$ remains larger than unity.

\begin{figure}
\onefigure{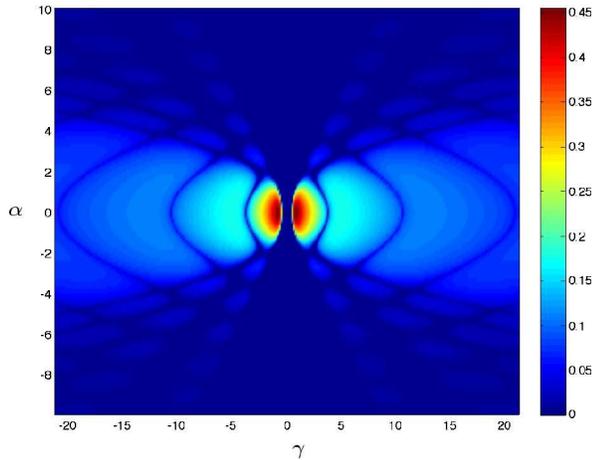}
\caption{(Color online) Growth rate $\sigma''\equiv\sigma'(\alpha, \gamma) /|\gamma|$ in the
  $(\gamma,\alpha)$--plane. Maxima are achieved at finite values of $\gamma$.
}
\end{figure}

\noindent {\bf Interpretation.}\! --
It should be stressed that the instabilities reported here are only made 
possible by the non self-adjoint nature of the $\calL$ operator, i.e.~of the induction equation.
If the system was governed by a negative self-adjoint operator, it could
 be readily
shown that a time dependence of the form (\ref{system_puls}) would 
imply negative values of
\begin{equation}
\tfrac{1}{2}\, \dR_t |\bfB|^2 = {\langle \calL \RC^{-1}(\theta (t))
\, \bfB \,|\, \RC^{-1}(\theta (t))
\, \bfB \rangle \, \leq 0 \, .}
\end{equation}

The role of transient growth in this 
parametric instability can be further analysed by taking advantage of the 
fact $\beta \rightarrow 0$. {The system can then be investigated under 
the quasi-static approximation of constant $\dot{\theta} = \alpha_0$,} and
locally replaced near each time 
 $t_0$ by \begin{equation}
\LC(t) = \RC(\alpha_0 \, t) \,\, \calLU \,\,\RC(\alpha_0 \, t)^{-1} \,,
\label{system_spirale}
\end{equation} 
where  $\alpha _0 = \alpha  \beta \, \cos (\beta\, t_0)$.
This system corresponds to a steady rotation of the shearing flow at
angular velocity $\alpha_0$ {(such situation can occur in convectively
driven dynamos, e.g. \cite{Soward3}).}
It is completely integrable, and admits exponentially growing
solutions for $ -1 < \alpha_0 < 0\, ,$
\begin{equation}
\sigma = -\varepsilon + \sqrt{-\alpha_0 \, (1+\alpha_0)} \, .
\label{cercle}
\end{equation}
{This formula is reminiscent of the classical 
Rayleigh-Pedley dispersion relation for rotating shear flows,
in which $\alpha_0$ is replaced by $2 \, \alpha_0$
\cite{Pedley}. This similarity is due to the 
well known formal analogy of the induction 
and vorticity equations. The governing pertubation 
equations do however differ.}

This exponential growth can be directly related to the transient 
field amplification in the initial system (\ref{goveq}). While an initial
magnetic field is amplified by the shear as described by 
(\ref{transientgrowth}), its direction is altered. This results
in a decreases of the phase with time.
Let us consider this mechanism over a small increment of time $\delta t$. 
The initial field is amplified 
\begin{equation}
||\bfB(t+\delta t)||=\rme ^{\hat \sigma \delta t}\, ||\bfB(t)||\, ,
\end{equation}
and rotated by a small angle
\begin{equation}
\bfB(t+\delta t)/||\bfB(t+\delta t)||=\RC(\hat \alpha \delta t) 
\, \bfB(t)/||\bfB(t)|| \, .
\end{equation}
If we now consider a system of the form (\ref{system_spirale}), 
$\alpha_0$ can be set to $\hat \alpha$ so as to cancel the effect of rotation,
which leaves us with $\bfB(t+\delta t)=\rme ^{\hat \sigma \delta t} 
\, \bfB(t)$. 
Incrementing in time then yields an exponential increase of 
the amplitude with  growth rate  $\hat \sigma$.
Initially, the field which is the most amplified by (\ref{goveq}) 
is $(1,1)^t/\sqrt{2}$, and corresponds to
$\hat \sigma= \tfrac{1}{2}-\varepsilon $ and
$\hat \alpha = -\tfrac{1}{2}$.
This is precisely the maximum growth rate in (\ref{cercle}),
and establishes the direct connection between the 
change in the field orientation during transient amplification 
and the exponential growth obtained with (\ref{system_spirale}).
The rotation rate $\hat \alpha$ provides a typical time scale 
$\hat \alpha^{-1}$ for systems exhibiting algebraic growth. 
This time scale will provide the physical interpretation 
of the instability process investigated here. 

\begin{figure}
\onefigure{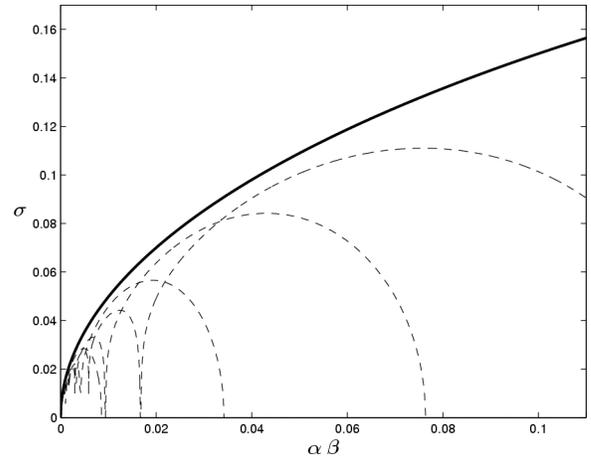}
\caption{Growth rate in the advective time scale versus $\alpha \,
\beta$. Dashed lines correspond to the result of a Floquet analysis on
system \eqref{system} at fixed values of $\alpha$, from right to left
respectively $\alpha = 0.25, 0.1875, 0.125, 0.0625$. The solid black curve 
corresponds to the estimate \eqref{racine}. }
\label{last_fig}
\end{figure}

If we return to (\ref{pulsations}) and use the approximation 
(\ref{system_spirale}), the rotation rate is provided by $\alpha \beta 
\cos (\beta t_0)$. For small enough $\alpha \beta$, formula \eqref{cercle}
can be used over half a period, corresponding to $\cos (\beta t_0) < 0$. 
This expression being related to the fastest growing mode,
the phase of the solution will act to reduce the realized 
growth rate. During the other half of the period $(\cos (\beta t_0) > 0)\, ,$ 
algebraic growth is achieved. A good approximation of the
resulting growth rate can be obtained by simply considering 
the growth rate provided by \eqref{cercle} and $\alpha_0=\alpha \,
\beta$ on the first half period 
and a zero growth rate for the reminder. As $\alpha \beta
\ll 1$ this  yields at leading order
\begin{equation}
\sigma = \tfrac{1}{2} \, \sqrt{\alpha \, \beta } \, .
\label{racine}
\end{equation}
Direct comparison of this expression 
(still on the advective time scale) with the Floquet
integration is provided on Figure~\ref{last_fig}.

Expression \eqref{cercle} also sheds light on the requirement $\varepsilon
\rightarrow 0$ for instability in the limit of infinitesimal perturbations.
Indeed $\alpha \beta > \varepsilon ^2$ is needed to allow positive values
of $\sigma$.

The exponential instabilities investigated here can therefore be understood
both in terms of parametric instabilities and in terms of perturbed
algebraic amplification. It is interesting to note that the steady
operator  
$\calL$ is not an oscillator and has no characteristic frequency. Since it 
is associated with transient amplification, a typical time scale $\hat \alpha
^{-1}$ can, however, be defined as introduced above. 
The evolution of the unperturbed system provides, as for the pendulum
(Mathieu equation),
the natural time scale for parametric amplification.

\noindent {\bf Conclusions.}\! --
In the limit of large magnetic Reynolds number $\Reym = \varepsilon^{-1}$, a
distinction is usually made between dynamo action on the ``slow'' diffusive
time scale $\tau_\eta$ and the ``fast'' advective time scale
$\tauU$. We stress here the potential importance of a
third time scale corresponding to flow modulations (here $\tau_\beta$), on
which exponential growth can be achieved.
Natural dynamos often exhibit large scale shear (usually in the form
of a differential rotation associated with the so-called
``$\Omega$--effect''). Such large scale shear are unable to produce
dynamo action on their own. Although the model we present is extremely
simplified (the shear is uniform and the field only depends on one
spatial coordinate), we believe the conclusion that interaction of shear
and rapid fluctuations can result in dynamo action is an important property.
Although we restricted our attention to a simplified setup,
interesting conclusions can be drawn concerning dynamo action. In the
limit of large 
$\Reym$, the flow needs to be measured with a precision $\Reym^{-1}$ for its 
dynamo property to be determined. Tiny fluctuations (as small as $\Reym^{-1}$)
 can drastically change the dynamo properties of the flow.    
 In astrophysical objects $\Reym^{-1}$ is often as small
as $10^{-18}$, so tiny fluctuations (but larger than $\Reym^{-1}$) can 
modify a non-dynamo flow to provide exponential growth. 
Further studies are obviously needed to assess under which conditions
this amplifications is realised in fully three dimensional configurations.

The possibility to drive a dynamo through periodic modifications of
the flow has already been pointed out, but such an approach
was so far restricted to finite amplitude modulations.
We find that even small amplitude oscillations can act to drive parametric
instabilities, while the unperturbed operator is stable. 
More generally, many physical problems are
governed by non self-adjoint operators. The system we investigated being 
very simplified, it could be interpreted in a much more
general framework than that of the induction equation
{(for instance in hydrodynamics, see \cite{Zahn,Schmid}).}


\begin{thebibliography}{0}






\bibitem{Glatz} Glatzmaier G.A. and  Roberts P.H., 
{\it Nature}, {\bf 377}  (1995) 203.

\bibitem{Christ}  Christensen U.R., Olson P.,  Glatzmaier G.A., 
{\it Geophys. J. Int.}, {\bf 138}  (1999) 393.

\bibitem{Fast1} S. Childress and A.D. Gilbert,
{\it Stretch, Twist, Fold: The Fast Dynamo},
Springer-Verlag (Berlin), 1995.

\bibitem{Fast2}  Galloway D.J. and  Proctor M.R.E., 
{\it Nature}, {\bf 356}  (1992) 691.

\bibitem{Backus} 
Backus G., 
{\it Ann. Phys.}, {\bf 4} 
(1958) 372.

\bibitem{Soward}
Soward A.M.,
{\it Geophysical and Astrophysical Fluid Dynamics}, {\bf 73}
(1993) 179.

\bibitem{Soward2}
Soward A.M., {\it Physica D}, {\bf 76}  (1994) 181.

\bibitem{Gog} {Gog J.R., Oprea I., Proctor M.R.E. and Rucklidge A.M.,
{\it Proc. Roy. Soc. A}, {\bf 455} (1999) 4205.}

\bibitem{PF} {P\'etr\'elis F. and Fauve S., {\it Europhys. Lett.},
{\bf 76}-4 (2006) 602.}

\bibitem{Peyrot} {Peyrot M., Plunian F. and Normand C, {\it Phys. Fluids},
  {\bf 19} (2007) 054109.}

\bibitem{induction}
Roberts P.H., 
{\it J. Math. Anal. App.}, {\bf 1} 
(1960) 195.

\bibitem{Farrell} {Farrell B., Ioannou P.J., {\it Astrophys. J.}, {\bf 522} 
(1999) 1079.}

\bibitem{induction2} Gerard-Varet D., 
{\it Phys. Fluids A}, {\bf 14}-4 (2002) 1458.

\bibitem{induction3} Livermore P. and   Jackson A., 
{\it Proc. Roy. Soc.  A},  {\bf 460}  (2004) 1453.

\bibitem{Schmid} Schmid P. \&   Henningson D.,
{\it Stability and transition in shear flows},
Applied Mathematical Sciences (142),
{Springer-Verlag},
{New York},
2001.

\bibitem{Zeldovich}  Zeldovich Ya.,
{\it JETP}, {\bf 31}, 
(1956) 154
[Sov. Phys. JETP, {\bf 4} 
(1957) 460].

\bibitem{Zeldovich2}
Zeldovich Ya. and
 Ruzmaikin A.A.,
{\it JETP}, {\bf 51} (1980) 493.

\bibitem{Morse}  Morse P.M. \&   Feshbach H.,
{\it Methods of Theoretical Physics, Part I.}, 
New York: McGraw-Hill, pp. 556-557, 1953.

\bibitem{Pedley} {Pedley T.J., {\it J. Fluid Mech.}, {\bf 35}, (1969) 97.}

\bibitem{Soward3} {Soward A.M., {\it Phil. Trans. A}, {\bf 275}-1256 
(1972) 611.}

\bibitem{Zahn} {Zahn, J.P., Toomre J., Spiegel E.A. \& Gough D.O.,
{\it J. Fluid Mech.}, {\bf 64}, (1974) 319.}




\end{thebibliography}
\end{document}